\documentclass{iopconfser}
\usepackage{graphicx}
\usepackage{amsmath}

\begin{document}

\title{The Barrel Imaging Calorimeter of the ePIC detector at the Electron-Ion Collider}

\author{Bobae Kim$^{1}$ on behalf of the ePIC Barrel Imaging Calorimeter (BIC) Detector Subsystem Collaboration}
\affil{$^1$Physics Division, Argonne National Laboratory, Lemont, IL, USA}
\email{bobae.kim@anl.gov}

\begin{abstract}
The Electron-Ion Collider (EIC) is a next-generation facility expected to address fundamental open questions in nuclear physics, including the structure and dynamics of gluons in nucleons and nuclei. The Barrel Imaging Calorimeter (BIC) is the electromagnetic calorimeter of the ePIC detector, designed to meet the stringent
performance requirements of the EIC, providing precise measurements and particle identification of scattered electrons, photons, and neutral mesons.

The BIC is a high-granularity sampling calorimeter based on lead absorbers interleaved with scintillating fibers, complemented by pixelated imaging layers based on the AstroPix monolithic active pixel sensor. This hybrid design enables both accurate energy and position measurements, as well as three-dimensional shower
reconstruction, making it a key component for realizing the full physics potential of the EIC.

The BIC has successfully completed its Preliminary Design Review last year and is currently progressing toward the Final Design Review. In this talk, I will present an overview of the BIC concept and focus on its current status and performance results from bench tests, system integration studies, and recent beam tests.
\end{abstract}

\section{Introduction}
\begin{figure}[htbp!]
    \centering
    \includegraphics[width=0.5\linewidth]{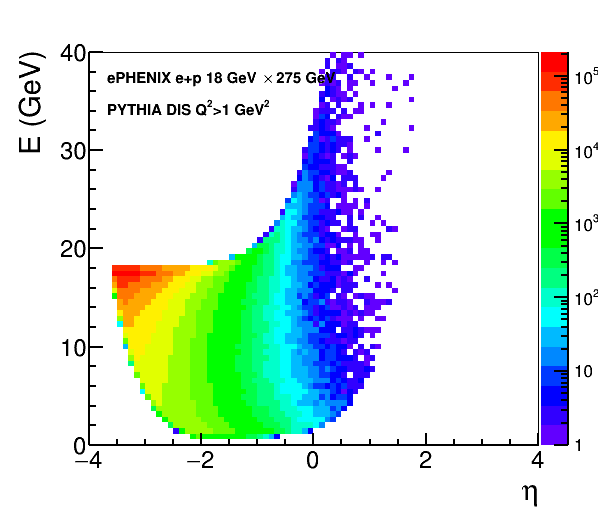}
    \includegraphics[width=0.47\linewidth]{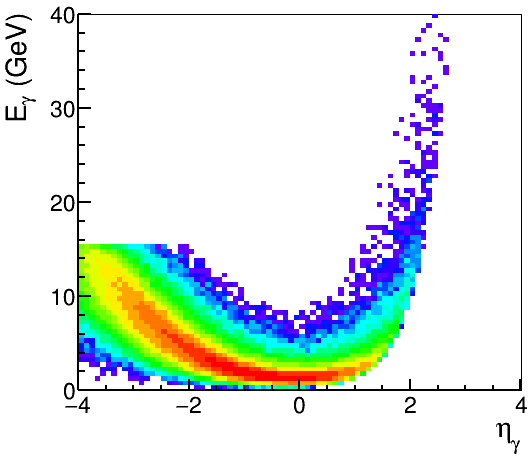}
    \caption{Calculated momentum spectra of particles in e+p 18$\times$275 GeV collisions: DIS $e^-$ from PYTHIA (left) and DVCS $\gamma$ from MILOU (right). Taken from Ref.~\cite{yp}.}
    \label{fig:2dplot}
\end{figure}

The Barrel Imaging Calorimeter (BIC) is one of the calorimeter subsystems of the ePIC detector at the Electron–Ion Collider (EIC), designed to measure electromagnetic showers in the barrel region.
It is designed to identify scattered electrons and measure their energies in high-$Q^2$ events, as well as to detect electrons and photons from particle decays and deeply virtual Compton scattering (DVCS) processes. Photon measurements and $\gamma/\pi^0$ discrimination further support neutral-meson identification.
Figure~\ref{fig:2dplot} shows the simulated momentum spectra of scattered electrons from deep-inelastic scattering (DIS) and photons from DVCS, generated with PYTHIA and MILOU, respectively, for collisions of 18 GeV electrons with 275 GeV protons~\cite{yp}.

According to the official ePIC Barrel EMCal System Requirements~\cite{epic_requirements}, the BIC must provide electron identification from below 1 GeV up to 50 GeV, with an energy resolution of $\sigma_E/E < 10\%/\sqrt{E} + (2\text{--}3)\%$. Additional requirements include e/$\pi$ separation down to 1 GeV/$c$, photon measurements and $\gamma/\pi^0$ discrimination up to 10 GeV, and the capability to distinguish two showers with an opening angle as small as 30 mrad. The system must also have sufficient dynamic range to detect minimum-ionizing particle (MIP) signals in all layers.

\begin{figure}[htbp!]
    \centering
    \includegraphics[width=0.75\linewidth]{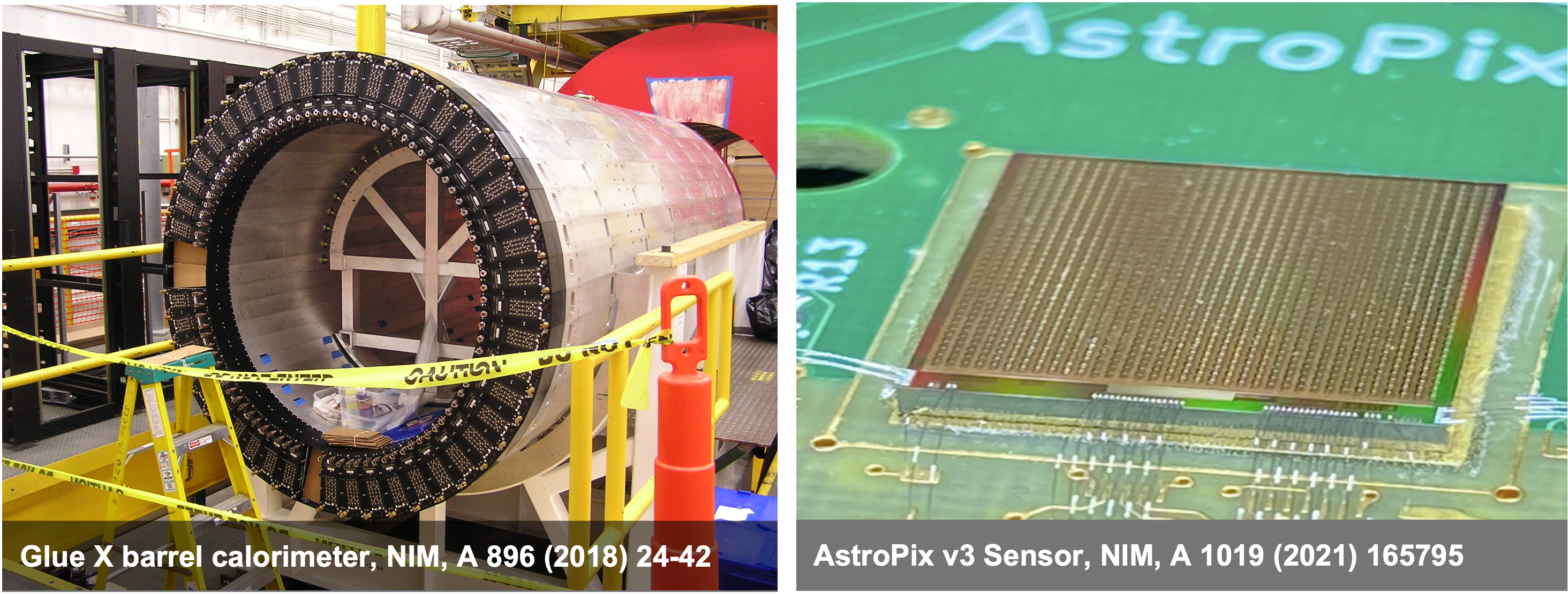}
    \caption{BIC concept: the GlueX barrel calorimeter~\cite{gluex2018} (left) and an AstroPix v3 sensor~\cite{astro2021} (right).}
    \label{fig:biccon}
\end{figure}

To meet these requirements, the BIC employs a hybrid calorimeter design combining a lead/scintillating-fiber (Pb/SciFi) sampling calorimeter with high-granularity silicon imaging layers based on AstroPix sensors, as shown in Figure~\ref{fig:biccon}.
The Pb/SciFi sampling calorimeter with two-sided SiPM readout is based on mature technology used in the GlueX and KLOE barrel calorimeters, with demonstrated detector performance. The energy and position resolutions at GlueX are $5.2\%/\sqrt{E} \oplus 3.5\%$~\cite{gluex2018} and 1.1 cm$/\sqrt{E}$ in the z-direction~\cite{gluex2008}, respectively.

The imaging layers use AstroPix sensors, which are high-voltage CMOS monolithic active pixel sensors (HV-CMOS MAPS) originally developed for NASA's AMEGO-X space mission~\cite{astro2021}. The $500~\mu\mathrm{m}$ pixel pitch corresponds to an expected intrinsic position resolution of approximately $144~\mu\mathrm{m}$ in the $z$-direction, complementing the coarser position resolution of the Pb/SciFi sampling calorimeter. The imaging layers are embedded in the first half of the calorimeter to capture three-dimensional images of developing electromagnetic showers. Additionally, AstroPix is designed to achieve an energy resolution better than 10\% (FWHM) at 122 keV and a dynamic range of 25--700 keV.

The baseline design incorporates four imaging layers, with mechanical provisions for up to six layers. The first imaging layer is also required to provide a charged-particle tracking point behind the DIRC detector with a spatial resolution better than 150 $\mu$m, supporting charged-hadron identification.

\subsection{Structure}
\begin{figure}[htbp]
    \centering
    \includegraphics[width=0.8\linewidth]{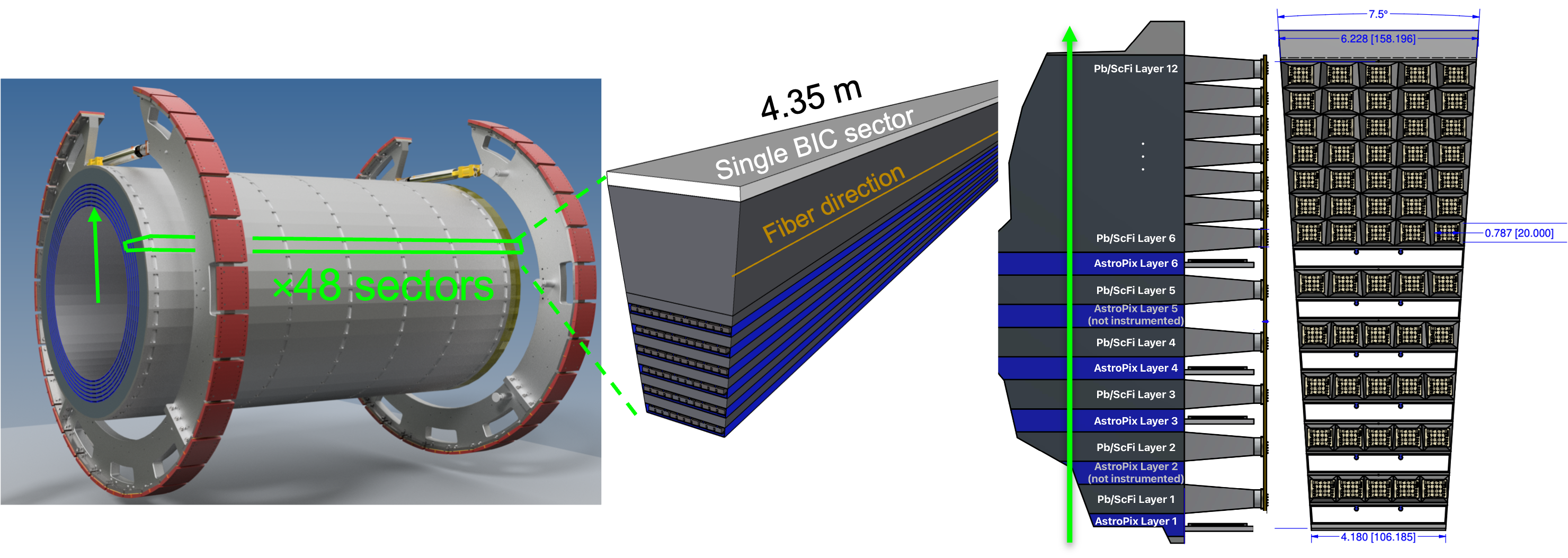}
    \caption{BIC detector structure: overall barrel geometry (left), longitudinal and cross-sectional views of a single sector (middle), and readout-cell segmentation of a Pb/SciFi layer (right).}
    \label{fig:bicstr}
\end{figure}
\begin{figure}[htbp]
    \centering
    \includegraphics[width=0.8\linewidth]{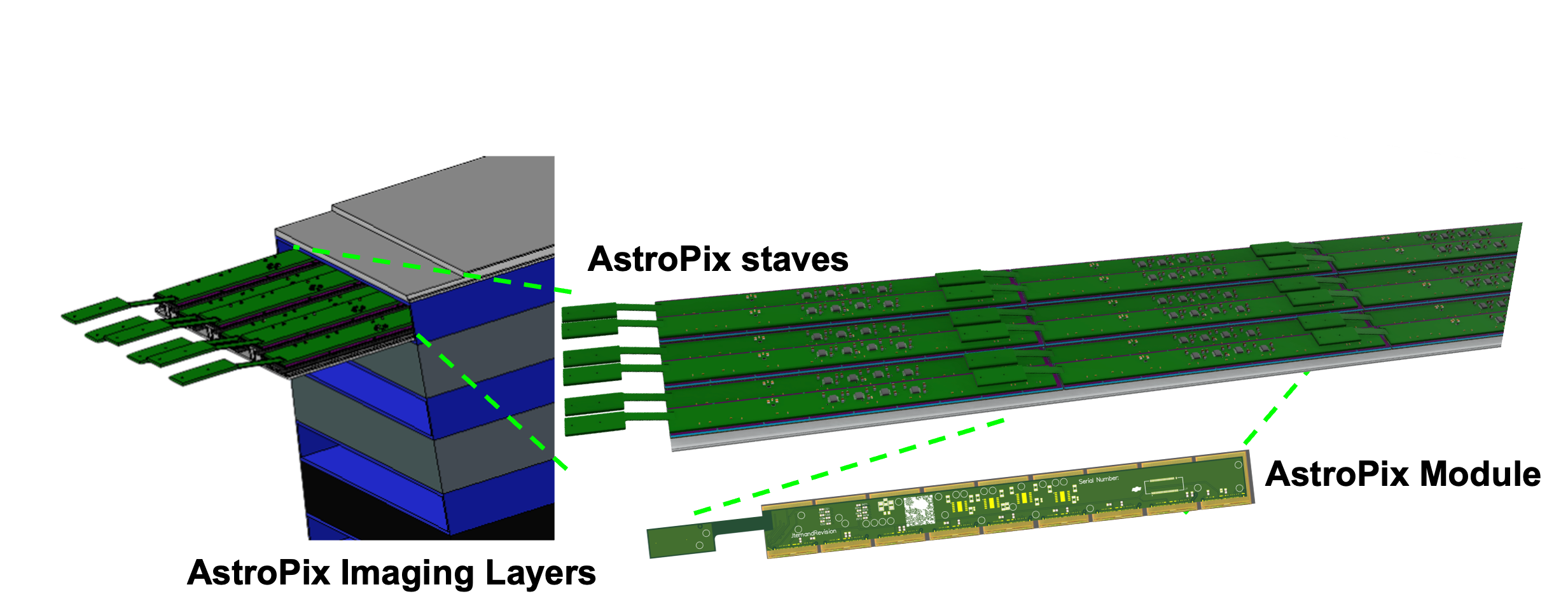}
    \caption{Structure of an AstroPix imaging layer.}
    \label{fig:bicimg}
\end{figure}

The BIC has an active length of 435 cm and an inner radius of 82 cm, covering a pseudorapidity range of $-1.71 < \eta < 1.31$. The barrel is divided into 48 sectors. The total depth exceeds 17.1 $X_0$ at $\eta = 0$, with a sampling fraction of approximately 10\%. It has six imaging-layer slots, four of which are populated with AstroPix imaging layers in the baseline design. These imaging layers are interleaved with five Pb/SciFi layers in the front section of the calorimeter, followed by a thicker rear Pb/SciFi section.

Each sector of the Pb/SciFi calorimeter consists of 12 sampling layers, each corresponding to approximately 1.4 $X_0$, as shown in Figure~\ref{fig:bicstr}. Each layer is divided into five readout cells, each equipped with individual light guides and SiPMs, resulting in a total of 120 SiPMs (Hamamatsu S14161-3050) per sector for two-sided readout. The SiPM signals are processed by CALOROC front-end electronics and transmitted to FELIX (Front-End Link eXchange), a PCIe-based data acquisition interface, via a common readout board (RDO).

Each AstroPix imaging layer consists of six to seven staves, as shown in Figure~\ref{fig:bicimg}. Each stave is 217.5 cm long, covering half of the BIC active length. Staves are installed from opposite ends of the barrel to provide full longitudinal coverage. Each stave comprises 12 modules arranged in series, and each module contains nine daisy-chained AstroPix chips mounted on an AstroLinx readout PCB. The data are transmitted via the End-of-Tray Card (ETC) to FELIX.

\section{Development Timeline}
\subsection{Pb/SciFi Calorimeter Layer}
\begin{figure}[htbp]
    \centering
    \includegraphics[width=0.87\linewidth]{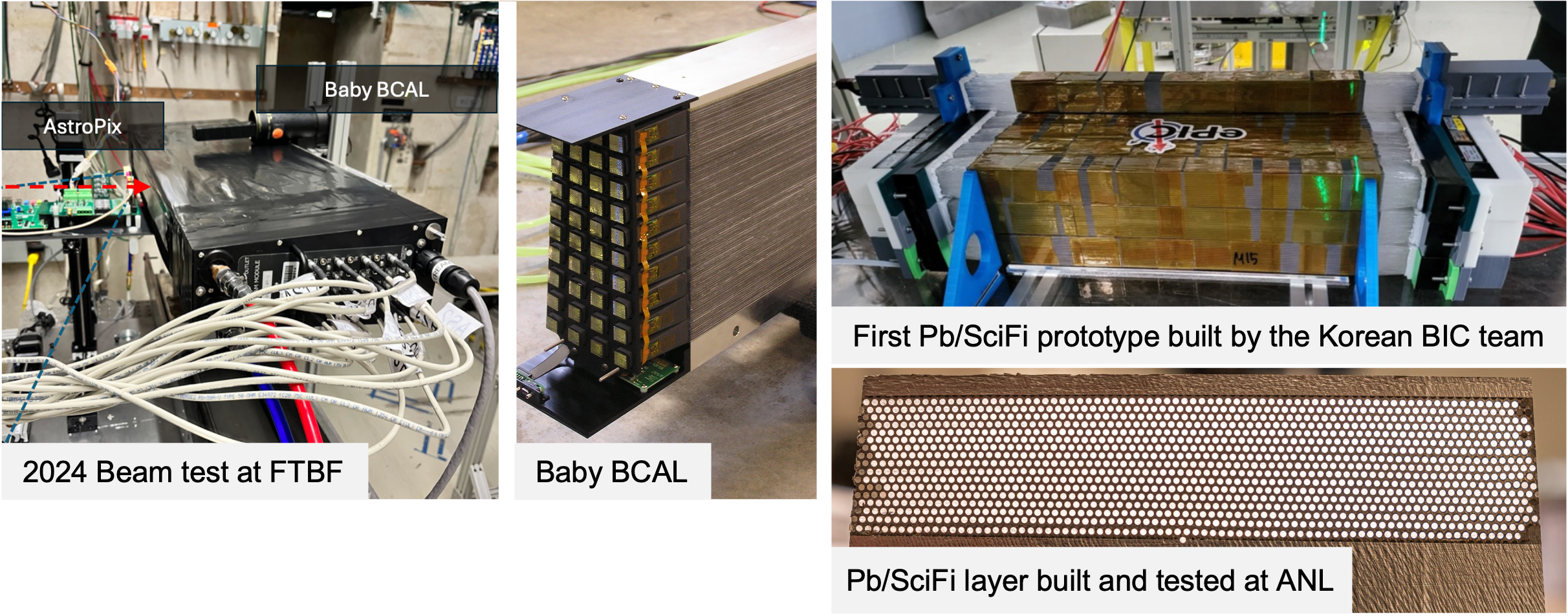}
    \caption{Pictures of fabricated Pb/SciFi prototype calorimeter layers.}
    \label{fig:pbscifi}
\end{figure}

The development of the Pb/SciFi calorimeter has progressed through several prototype construction and beam-test campaigns since 2024. In 2024, a prototype of the GlueX barrel calorimeter (Baby BCAL), with a transverse length of approximately 60 cm and a depth of 15.5 $X_0$, was tested at the Fermilab Test Beam Facility (FTBF). The prototype was equipped with 40 light guides and 40 SiPM arrays on each side. The beam-test results, including the energy response and e/$\pi$ separation performance, were published in JINST~\cite{Klest_2025}.

In parallel, the Korean BIC team constructed its first Pb/SciFi prototype in 2024, with a transverse size of 32 cm and a depth of 10.9 $X_0$, and increased its depth to 15 $X_0$ in 2025, although their configurations were not yet identical to the final BIC design. Beam tests were conducted at CERN and KEK during 2024--2025, and the results from the 2024 CERN beam test were reported in Ref.~\cite{lee2026beamtestpbscifiprototype}.

In 2025, Pb/SciFi sampling layers with a transverse length of 50 cm and a depth of 1.4 $X_0$ were fabricated at Argonne National Laboratory (ANL), using construction equipment originally developed for the GlueX BCAL at the University of Regina. Each layer was equipped with four light guides and four SiPMs per side. These layers were evaluated in the 2026 Hall D beam-test campaigns. 

In 2026, the Korean BIC team also fabricated Pb/SciFi layers with a transverse length of 70 cm using the same construction technique as that employed at ANL. A beam test at CERN was conducted for August 2026.
A full-scale 4.35-m-long sector is planned for preproduction in 2026, with subsequent calorimeter production planned at ANL and in Korea.

\subsection{AstroPix Imaging Layer}
\begin{figure}[htbp]
    \centering
    \includegraphics[width=1\linewidth]{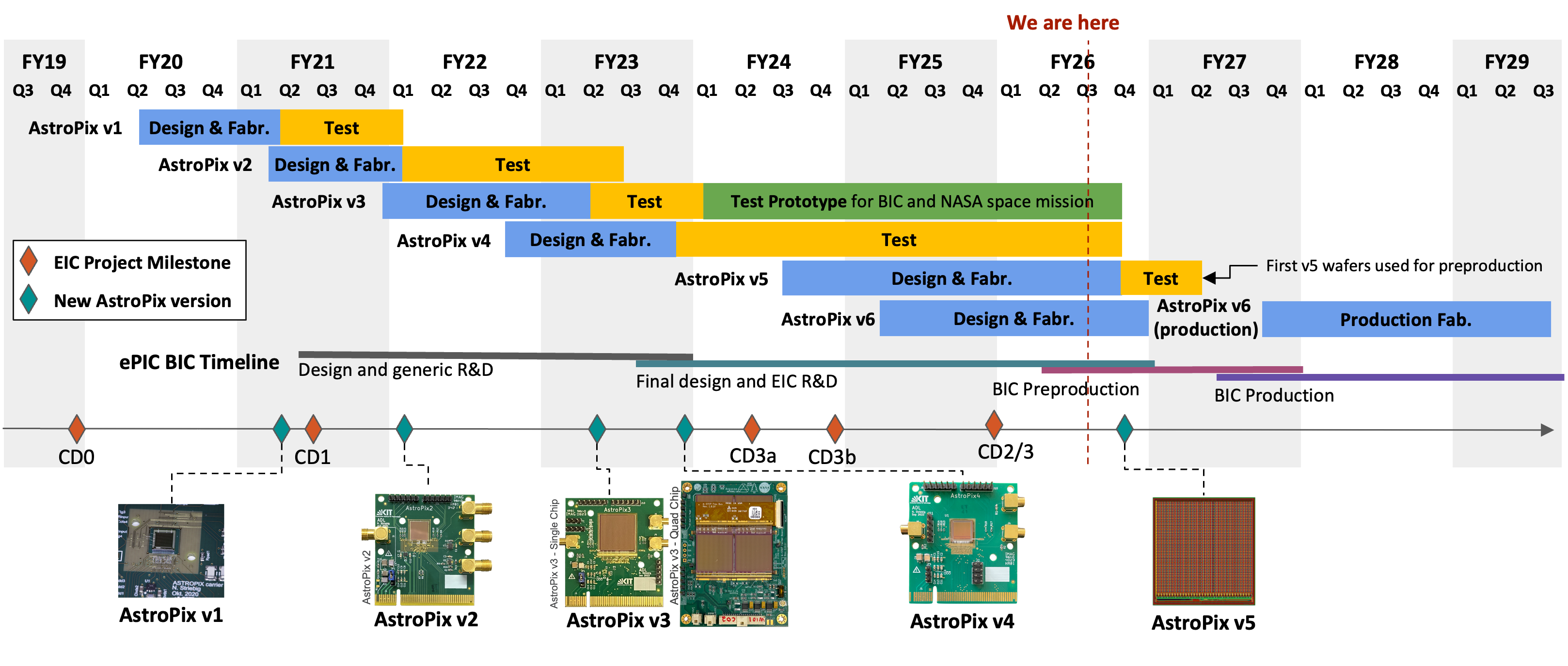}
    \caption{AstroPix sensor development timeline from v1 to v6, along with the ePIC BIC project milestones and production schedule.}
    \label{fig:astrord}
\end{figure}

The development timeline of AstroPix sensors, from v1 to v6, together with the ePIC BIC project milestones, is shown in Figure~\ref{fig:astrord}. The target specifications for the final AstroPix chip in the BIC include a pixel pitch of $500 \times 500~\mu\mathrm{m}^2$, an energy resolution better than 10\% (FWHM) at 122 keV, a dynamic range of 25--700 keV, and a fully depleted silicon thickness of 500 $\mu$m. A power consumption of approximately 2 mW/cm$^2$ and a time resolution of 20 ns are considered acceptable for AstroPix v5.

AstroPix v3 is the first full-size chip, featuring a $35 \times 35$ pixel matrix with a pixel pitch of $500~\mu$m~\cite{steinhebel2025astropix}. Single-chip characterization studies of v3 and v4 have been conducted, including energy calibration, MIP response, and the corresponding effective depletion depths. In particular, AstroPix v3 was tested with a 120 GeV proton beam at FTBF in June 2024, and its effective depletion depth was calculated from the measured MIP energy deposition~\cite{bobae2026astropixv3}. Similar characterization studies have also been performed with AstroPix v4, and the results are being prepared for publication. These studies help assess whether AstroPix can meet the BIC requirement to detect MIP signals in all imaging layers. The pre-production AstroPix v5 chip has been fabricated, with testing planned for January 2027.

\begin{figure}[htbp]
    \centering
    \includegraphics[width=0.8\linewidth]{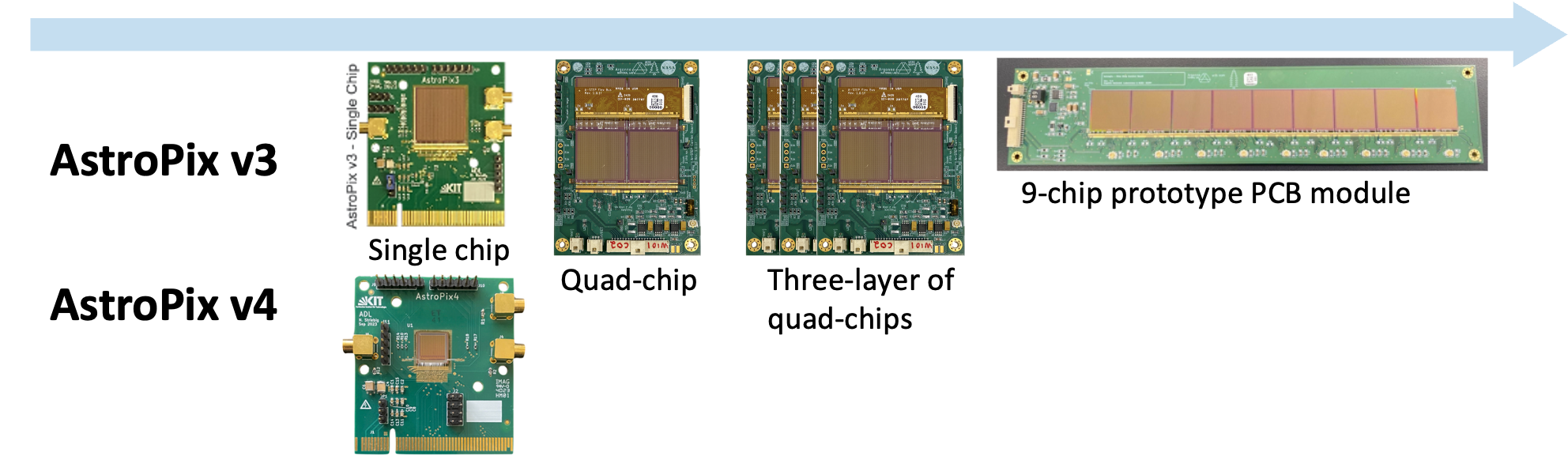}
    \caption{AstroPix imaging layer R\&D Milestones: from single-chip characterization to multi-chip and multi-layer prototype development.}
    \label{fig:astroimag}
\end{figure}
In parallel, multi-chip and multi-layer prototypes have been developed to demonstrate the scalability of the AstroPix imaging system, as illustrated in Figure~\ref{fig:astroimag}. A three-layer stack of AstroPix v3 single chips, with and without interleaved thin Pb/SciFi layers, was tested at CERN and KEK in 2025 using GECCO boards and a customized DAQ system. These tests aimed to measure chip-level hit efficiency and assess the ability of AstroPix to characterize electromagnetic shower development.~\cite{hong2026testbeamperformanceastropixsilicon}. 

Subsequently, similar studies were conducted using quad-chip prototypes instead of single chips. A quad-chip ($2 \times 2$) and a three-layer stack of quad-chips were tested at KEK in December 2025 using the A-STEP FPGA-based readout system. A two-layer stack of quad-chips interleaved with Pb/SciFi layers was subsequently tested in Hall D at Jefferson Lab in spring 2026. Finally, a nine-chip daisy-chained AstroPix v3 prototype module was tested at KEK in May 2026 and in Hall D at Jefferson Lab in August 2026. The module used a mock-up of AstroLinx, the readout board designed for the final BIC AstroPix modules. These tests aimed to demonstrate the feasibility of the BIC module readout design and evaluate multi-chip operation and beam-rate capability.

\subsection{System Integration}
Synchronization between the Pb/SciFi calorimeter and AstroPix readout systems is essential for integrated event building. Initial integration tests were conducted at FTBF in 2024 using the AstroPix analog signal as a trigger input to the JLab CODA-based DAQ system. Subsequent tests at KEK and CERN in 2025 used the AstroPix MISO signal as a trigger input.

The ultimate goal is to establish a common time base by distributing an external clock to the AstroPix and calorimeter readout systems, enabling synchronized event building through FPGA timestamp comparisons across the two different DAQ systems. Integration tests using the A-STEP and H2GCROC readout systems were conducted under beam conditions in Hall D at Jefferson Lab in 2026. During these tests, a common 40 MHz external clock was distributed to both readout systems. Analysis of the recorded FPGA timestamps is ongoing to evaluate event synchronization.

\section{BIC Beam Test Results}
\subsection{Pb/SciFi Prototype Performance}
The performance of the Baby BCAL prototype, including electron and pion responses and e/$\pi$ separation, was evaluated at the FTBF in 2024, and the results were published in JINST~\cite{Klest_2025}.

\begin{figure}[htbp]
    \centering
    \includegraphics[width=0.4\linewidth]{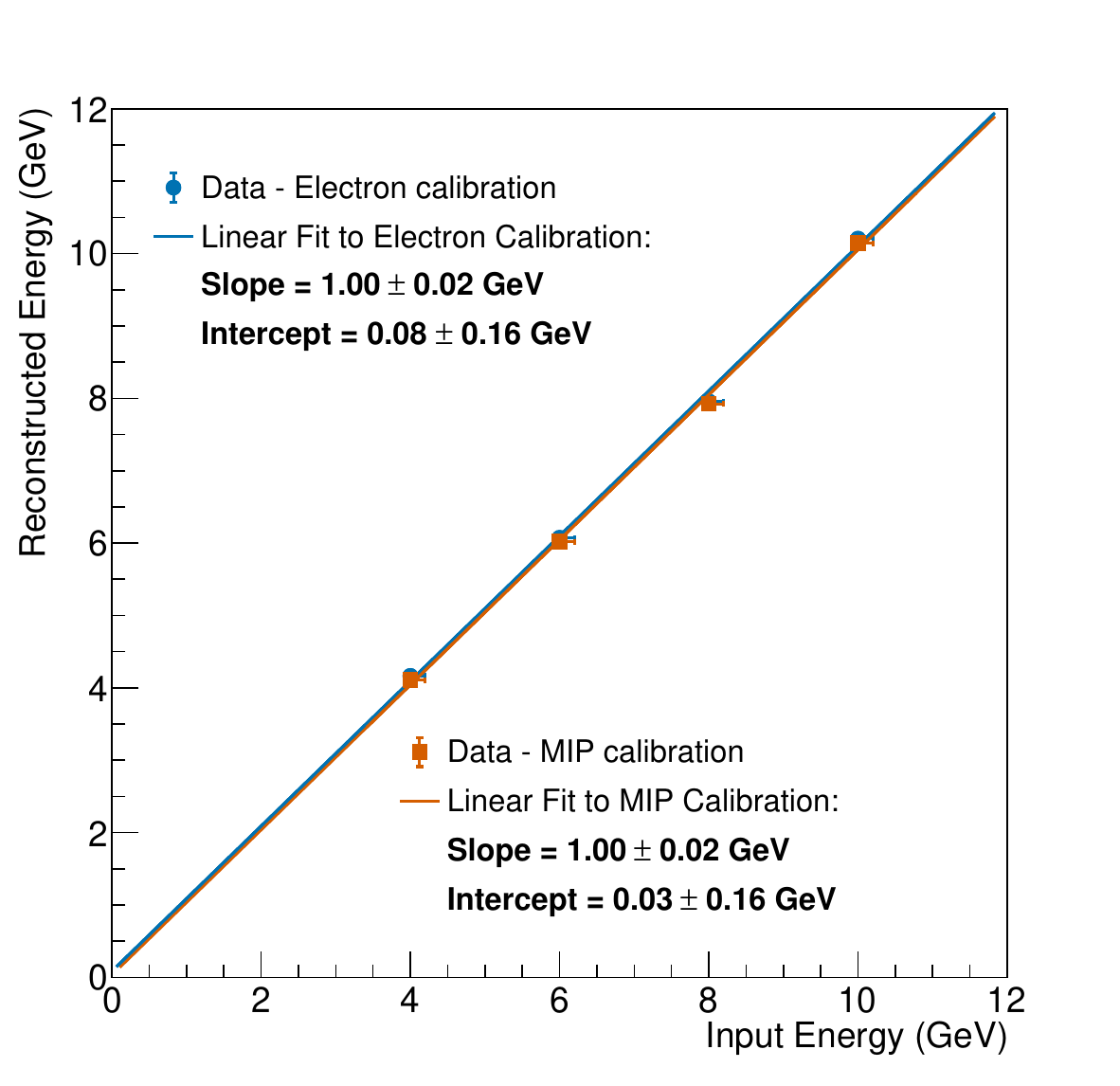}
    \includegraphics[width=0.5\linewidth]{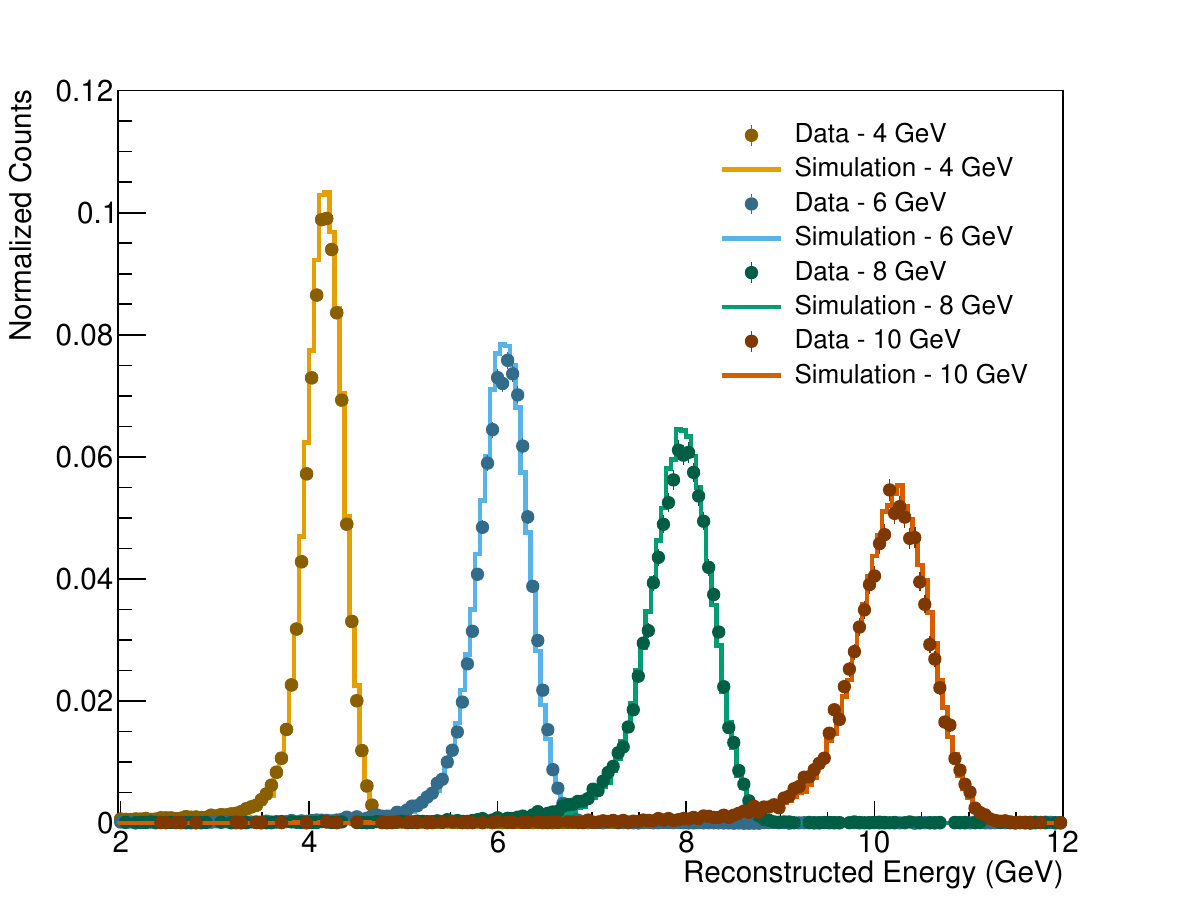}
    \caption{Linearity of the electron response (left) and comparison of electron responses between data and GEANT4 simulation (right)~\cite{Klest_2025}. Reproduced from Ref.~\cite{Klest_2025}.}
    \label{fig:bblin}
\end{figure}

\begin{figure}[htbp]
    \centering
    \includegraphics[width=0.48\linewidth]{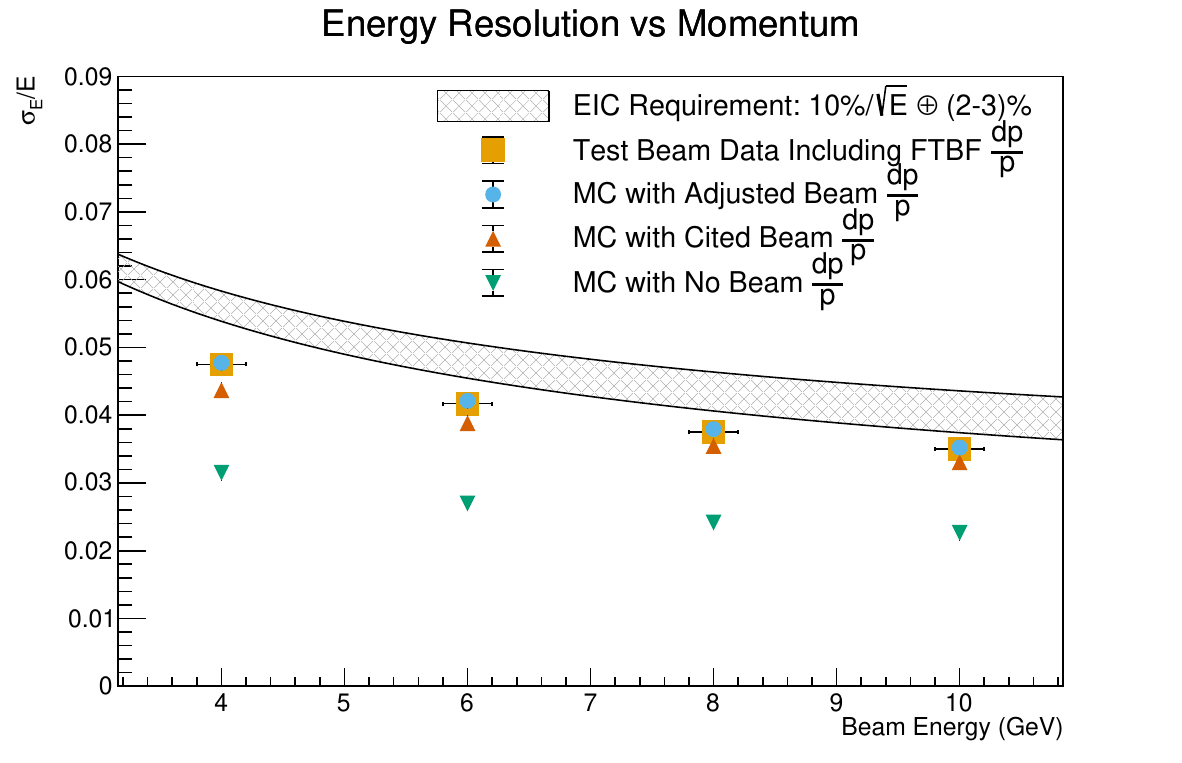}
    \includegraphics[width=0.45\linewidth]{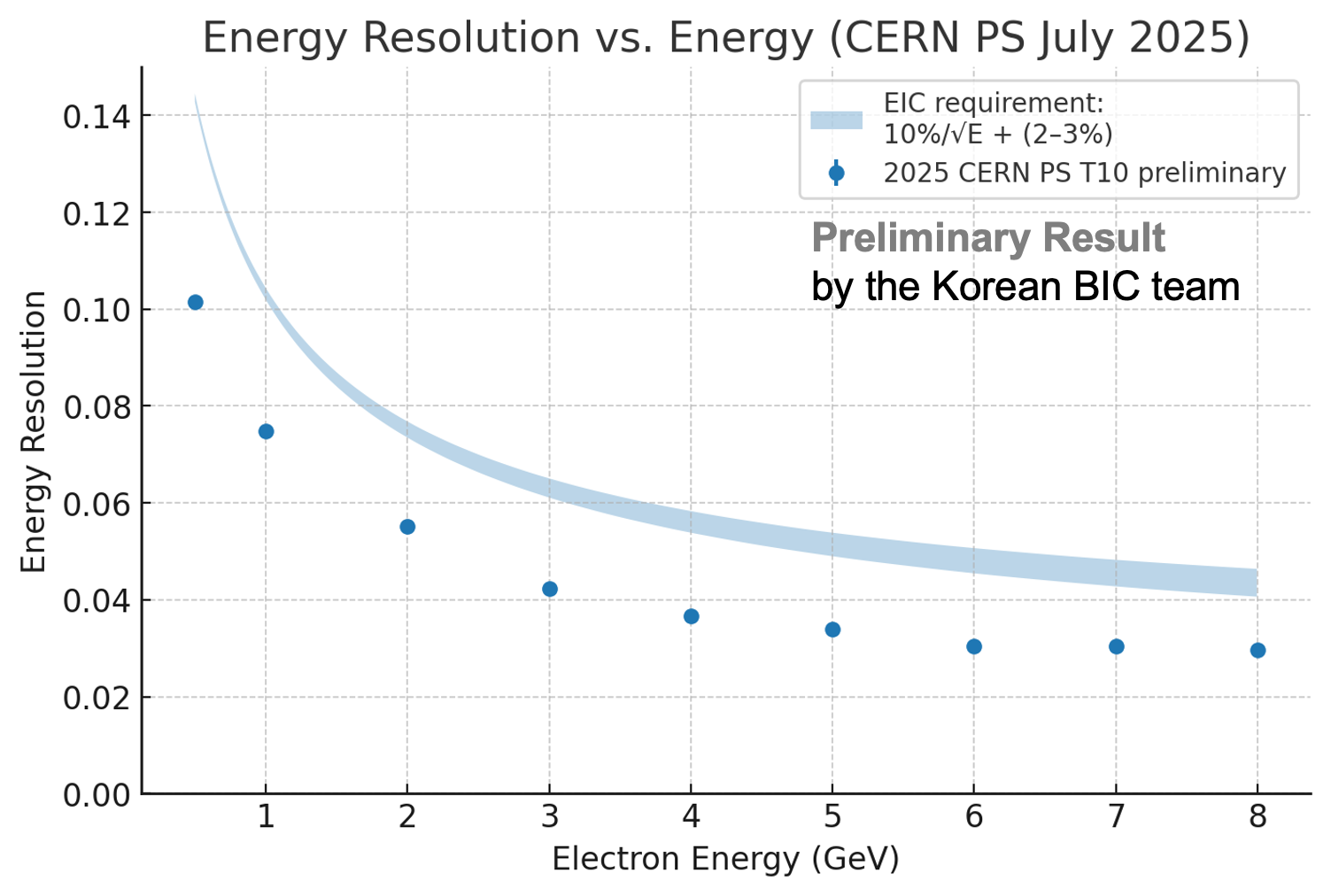}
    \caption{Electron energy resolution: Baby BCAL prototype at FTBF (left)~\cite{Klest_2025} and Pb/SciFi prototype developed by the Korean BIC team at CERN PS T10 (right).}
    \label{fig:bberes}
\end{figure}

The electron response of the 15.5 $X_0$ Baby BCAL prototype was evaluated using electron beams of 4, 6, 8, and 10 GeV. Two calibration methods based on MIPs and electrons were applied. The results demonstrated excellent linearity over the 4--10 GeV energy range and good agreement between data and GEANT4 simulation, as shown in Figure~\ref{fig:bblin}. Figure~\ref{fig:bberes} presents the measured electron energy resolutions, which satisfy the ePIC barrel EMCal system requirements. In addition, a 15 $X_0$ Pb/SciFi prototype with PMT readout was tested at CERN PS T10 in 2025, achieving an energy resolution that also meets the requirements, although the construction technique differed from that planned for the final BIC.

\begin{figure}[htbp]
    \centering
    \includegraphics[width=0.55\linewidth]{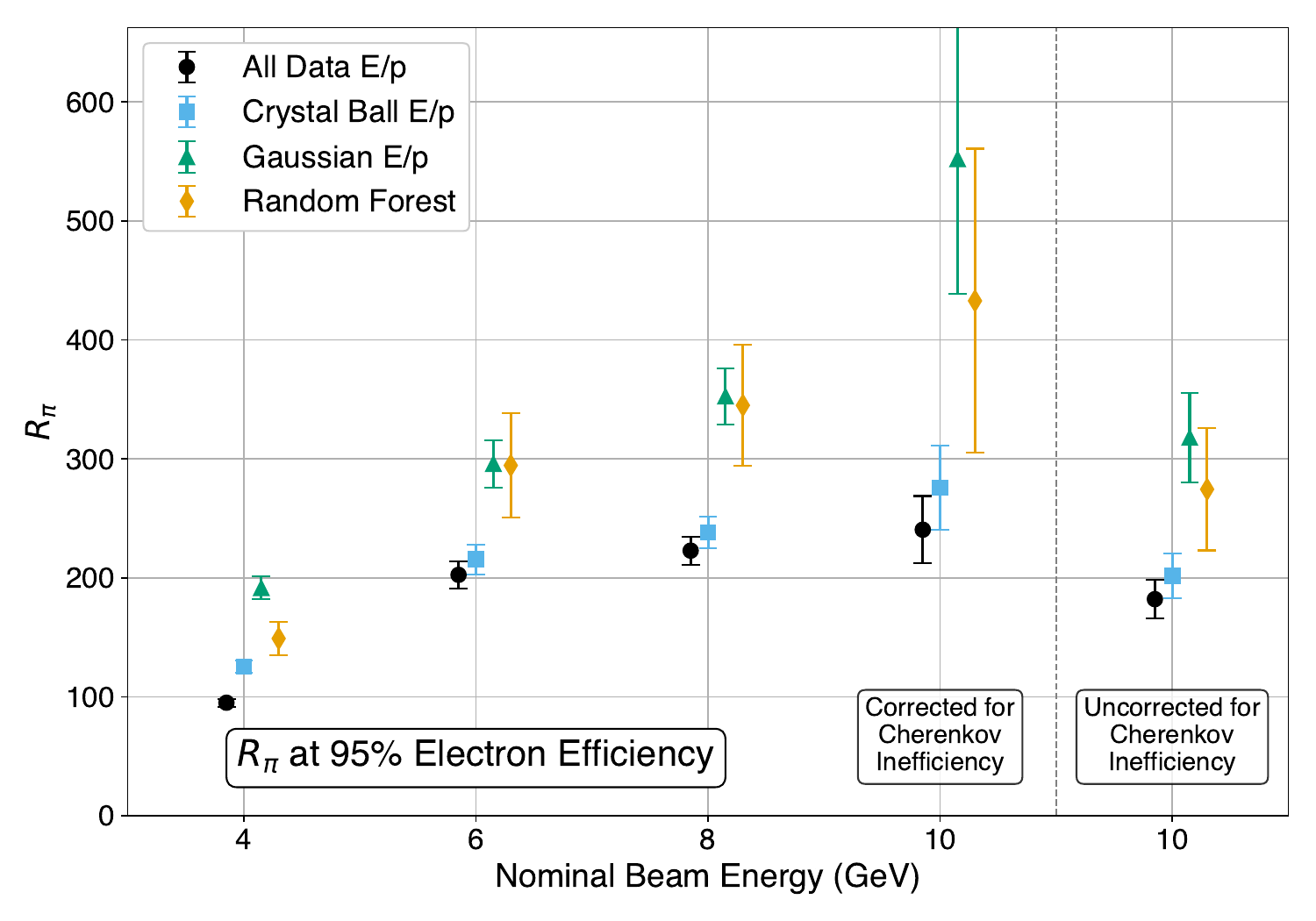}
    \caption{e/$\pi$ separation performance of the Baby BCAL prototype: pion rejection factor evaluated using $E/p$-based methods and a random forest classifier~\cite{Klest_2025}. Reproduced from Ref.~\cite{Klest_2025}.}
    \label{fig:bbsep}
\end{figure}

The pion response of the Baby BCAL was also studied using pion events selected by Cherenkov detectors at FTBF. The measured response was compared with GEANT4 simulations, showing good agreement at 6 GeV, while the simulation slightly overestimated the response at 4 GeV and underestimated it at 8 GeV. These differences may be partly attributed to uncertainties in the mean beam momentum and its spread. Overall, the simulation reproduced the measured pion response reasonably well, providing a benchmark for predicting e/$\pi$ separation performance in the BIC.

e/$\pi$ separation was evaluated using the Baby BCAL beam-test data and simulations. The pion rejection factor was estimated using three methods based on electron-peak modeling in the $E/p$ distributions, as well as a random forest classifier using ADC information from all readout channels. As shown in Figure~\ref{fig:bbsep}, at 95\% electron efficiency, the pion rejection factor ($R_\pi$) was approximately 100 or higher at 4 GeV and exceeded 200 at 6--10 GeV, with the 10 GeV results corrected for Cherenkov detector inefficiency. These results demonstrate high pion rejection factor using the Pb/SciFi calorimeter alone. Further improvement in e/$\pi$ discrimination is expected by combining the calorimeter response with high-granularity AstroPix hit information in the full BIC detector.

\subsection{AstroPix Imaging Layer Prototype Performance}
The performance of AstroPix prototypes was evaluated through bench tests and beam tests at several facilities, including the FTBF, KEK in Japan, and Hall D at Jefferson Lab. These studies focused on MIP response, shower imaging, and multi-chip and multi-layer operation to assess the suitability of AstroPix for the BIC imaging layers.

\begin{figure}[htbp]
    \centering
    \includegraphics[width=0.4\linewidth]{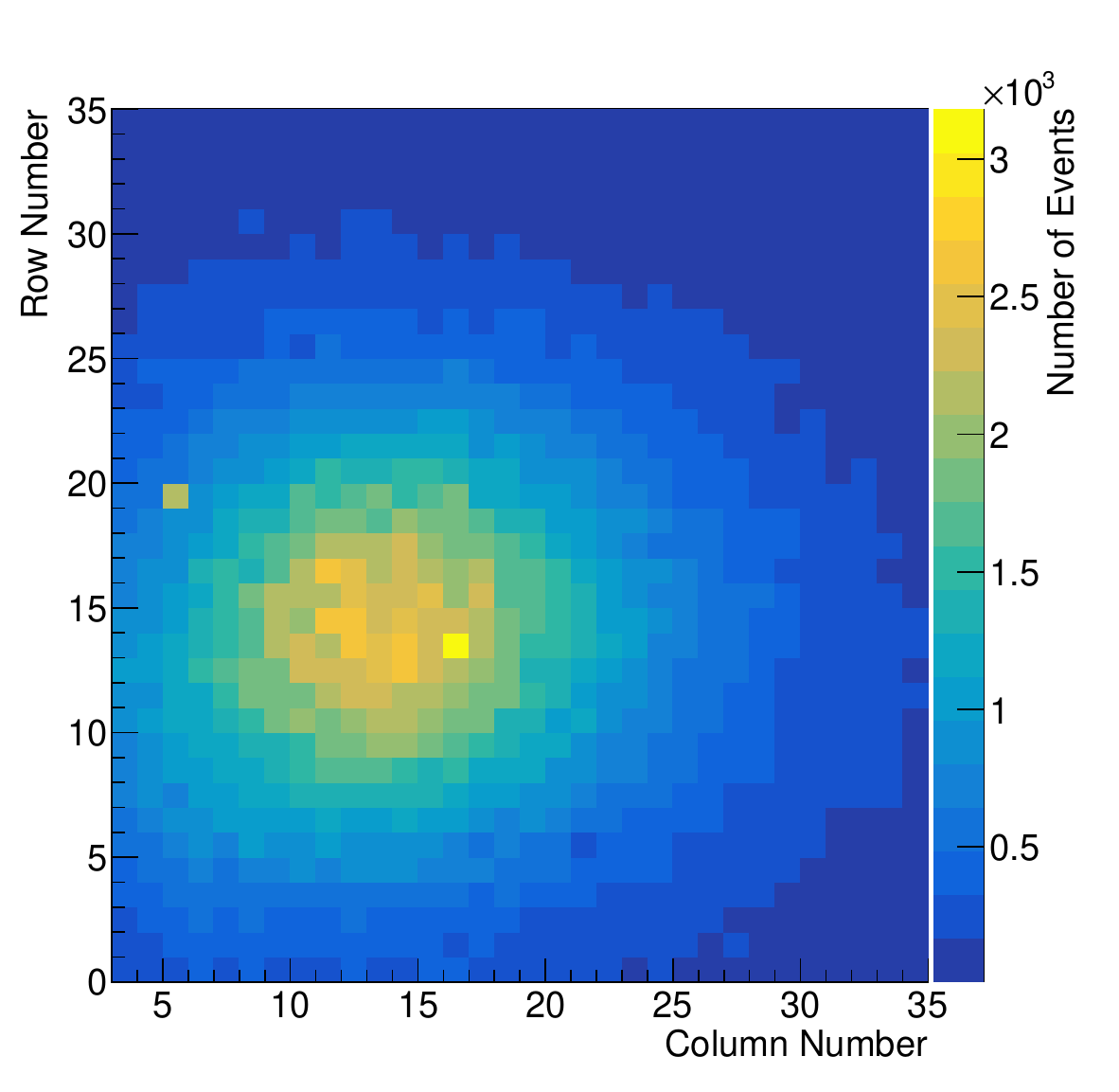}
    \includegraphics[width=0.5\linewidth]{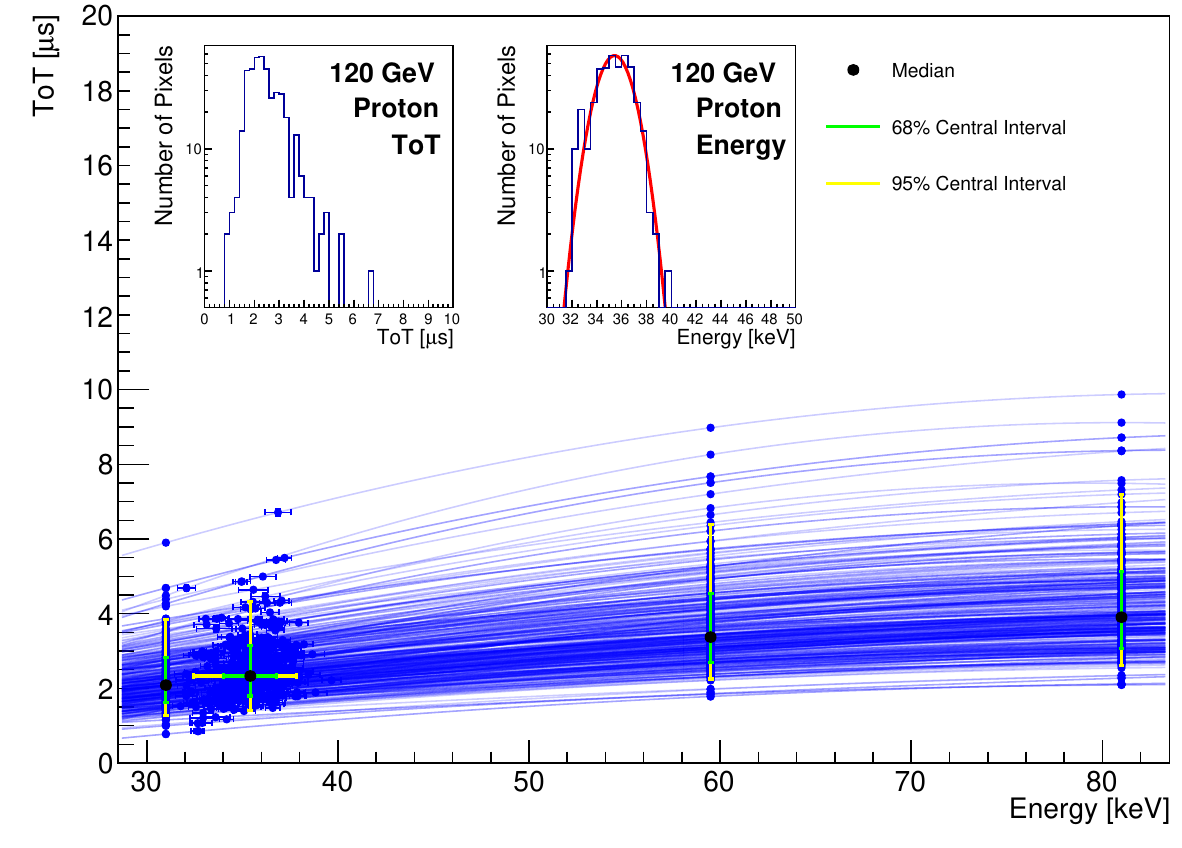}
    \caption{Example hit map obtained with 120 GeV protons (left) and per-pixel energy calibration curves overlaid with MPVs from the proton measurements (right). The upper subplots show the corresponding ToT and energy distributions, respectively~\cite{bobae2026astropixv3}. Reproduced from Ref.~\cite{bobae2026astropixv3}.}
    \label{fig:v3mip}
\end{figure}

The MIP response of an AstroPix v3 single chip was studied using a 120 GeV proton beam at FTBF~\cite{bobae2026astropixv3}. Figure~\ref{fig:v3mip} shows an example hit map obtained with a 120 GeV proton beam (left) and per-pixel energy calibration curves obtained using radioactive sources, overlaid with the most probable values (MPVs) extracted from the energy distributions of individual pixels (right). The measured MIP distributions are well described by a Landau function convolved with a Gaussian function. The MIP MPVs extracted for the selected pixels have a mean of 34 keV and a standard deviation of 1.9 keV, corresponding to an effective depletion depth of 128 $\pm$ 8 $\mu$m. This study provides the first characterization of the AstroPix v3 single-chip response to 120 GeV protons and demonstrates MIP detection within a sufficient dynamic range, satisfying one of the ePIC Barrel EMCal System Requirements.

\begin{figure}[htbp]
    \centering
    \includegraphics[width=0.48\linewidth]{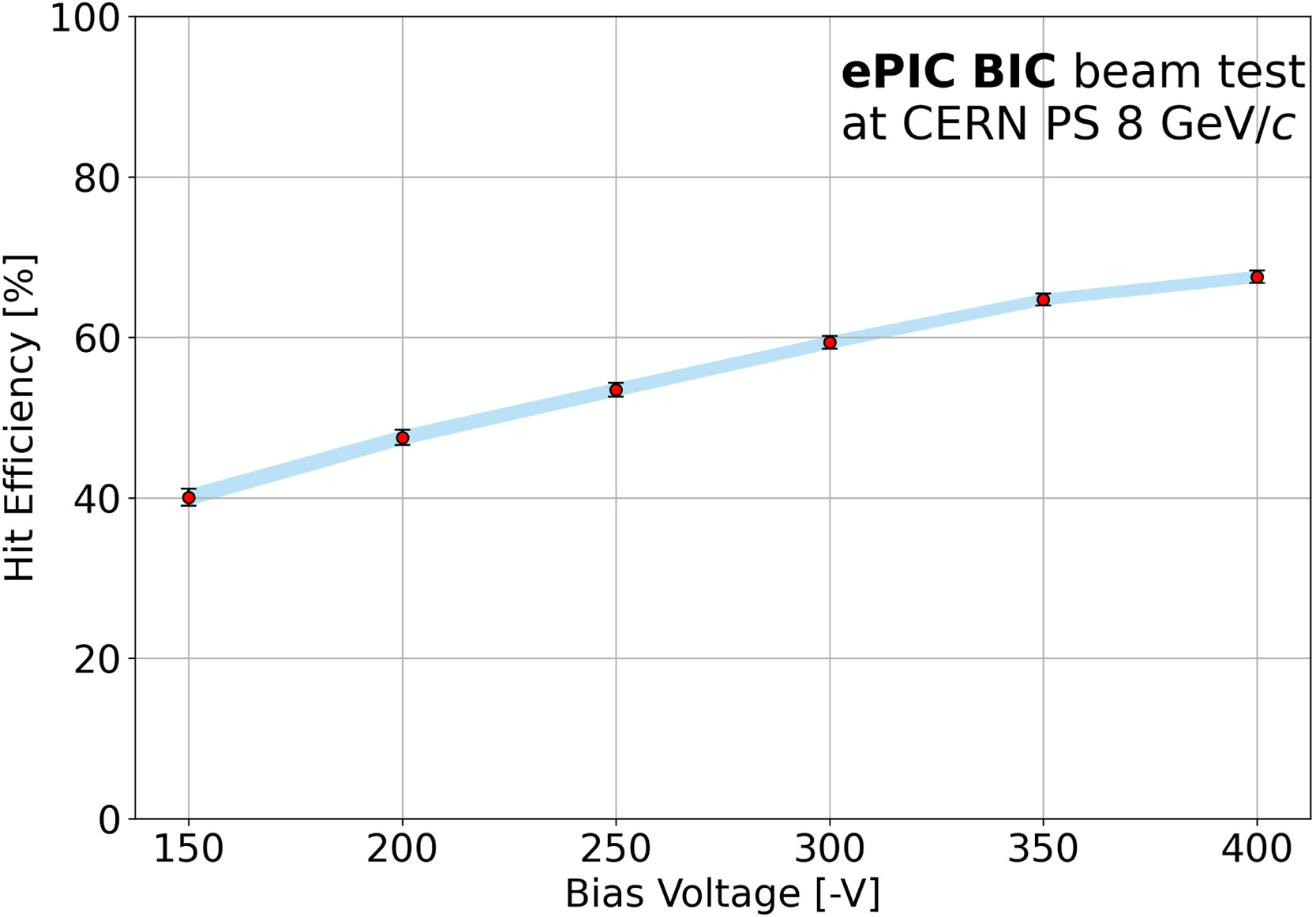}
    \includegraphics[width=0.45\linewidth]{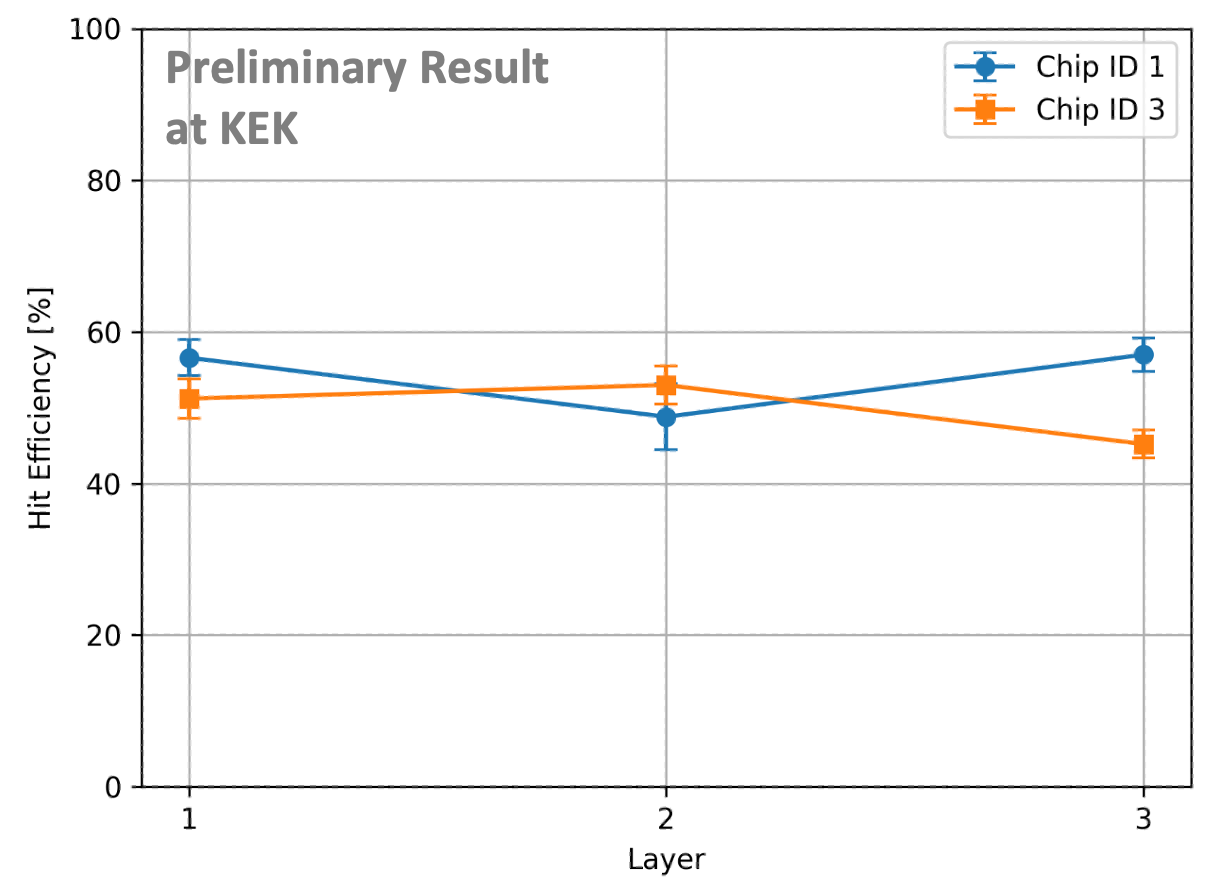}
    \caption{Measured hit efficiencies for three-layer configurations of AstroPix v3 single chips~\cite{hong2026testbeamperformanceastropixsilicon} (left) and quad-chips (right).}
    \label{fig:v3eff}
\end{figure}

Hit efficiencies were evaluated for two three-layer configurations: one using AstroPix v3 single chips~\cite{hong2026testbeamperformanceastropixsilicon} and the other using quad-chips, as shown in Figure~\ref{fig:v3eff}. For the single-chip configuration, the hit efficiency was measured as a function of bias voltage. For the quad-chip configuration, the hit efficiency of each layer was evaluated at a fixed bias voltage using data from chips with IDs 1 and 3, which had sufficient beam coverage and event statistics. AstroPix v3 exhibited hit efficiencies ranging from 40 to 68\%, with further improvements expected in AstroPix v5.

\begin{figure}[htbp]
    \centering
    \includegraphics[width=1\linewidth]{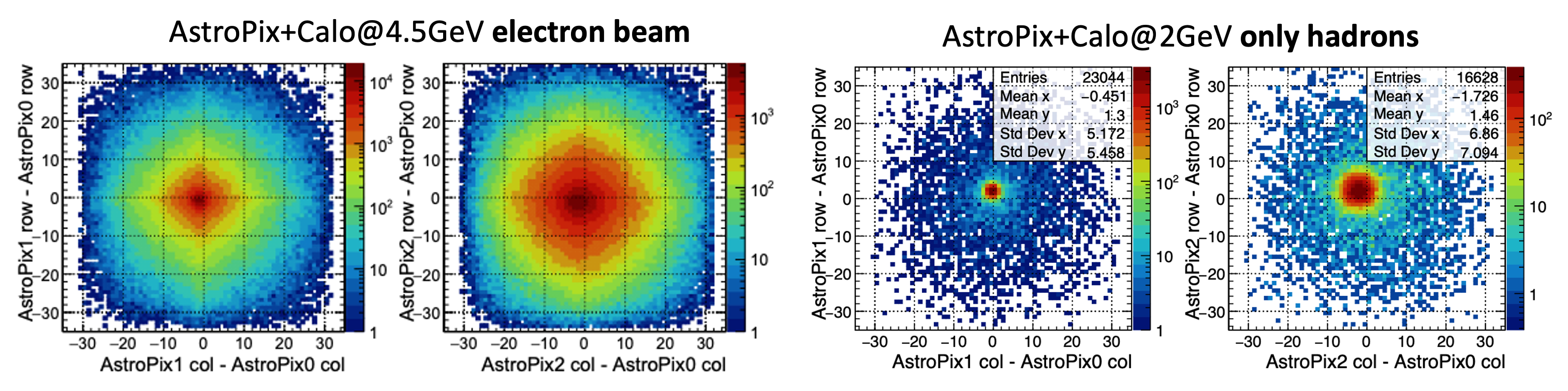}
    \caption{Position residual distributions obtained with 4.5 GeV electron (left two plots) and 2 GeV hadron (right two plots) beams using three AstroPix v3 single-chip layers~\cite{hong2026testbeamperformanceastropixsilicon}. Reproduced from Ref.~\cite{hong2026testbeamperformanceastropixsilicon}.}
    \label{fig:v3img}
\end{figure}

Three AstroPix v3 single-chip layers interleaved with Pb/SciFi modules were used to investigate the capability of AstroPix to characterize electromagnetic shower development using electron and hadron beams. Figure~\ref{fig:v3img} shows the position residual distributions (Layer 2--Layer 1 and Layer 3--Layer 1) obtained with 4.5 GeV electron (left two plots) and 2 GeV hadron (right two plots) beams. Distinct position residual patterns were observed for electrons and pions, suggesting their potential for e/$\pi$ discrimination.
The published study~\cite{hong2026testbeamperformanceastropixsilicon} provides a quantitative characterization of shower development using observables such as hit multiplicity and radial residuals, demonstrating the feasibility of shower imaging with small-scale prototype modules.

\begin{figure}[htbp]
    \centering
    \includegraphics[width=0.9\linewidth]{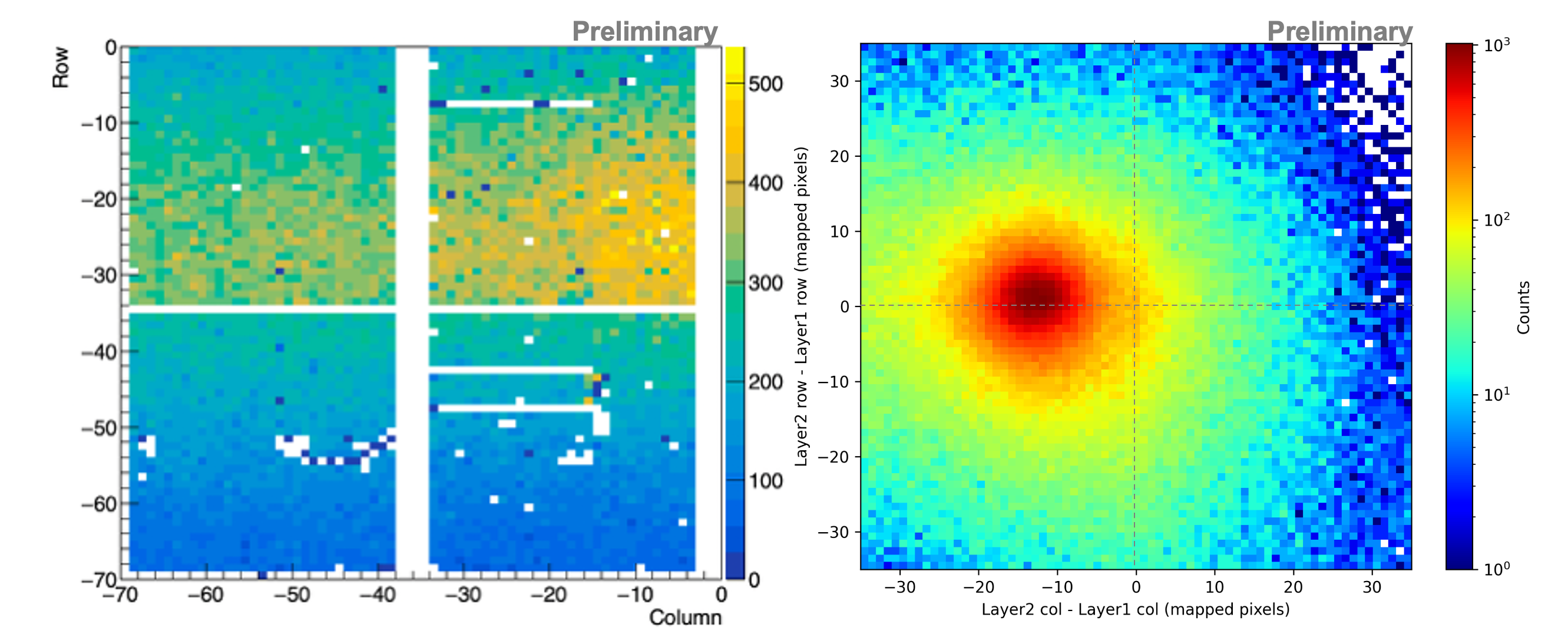}
    \caption{Example beam profile measured with the first AstroPix v3 quad-chip module (left) and position residual distribution (right).}
    \label{fig:v3imgq}
\end{figure}

A beam test using two AstroPix v3 quad-chip layers was performed with an electron beam in Hall D at Jefferson Lab. The configuration was limited to two quad-chip layers due to readout buffer limitations at high beam intensity. Figure~\ref{fig:v3imgq} shows the beam profile measured by the first AstroPix quad-chip and the position residual distribution (Layer 2--Layer 1) in the left and right plots, respectively. Two quad-chip layers exhibited a position residual pattern consistent with that observed in the single-chip measurements, demonstrating successful operation of the extended daisy-chained quad-chip configuration.

\begin{figure}[htbp]
    \centering
    \includegraphics[width=1\linewidth]{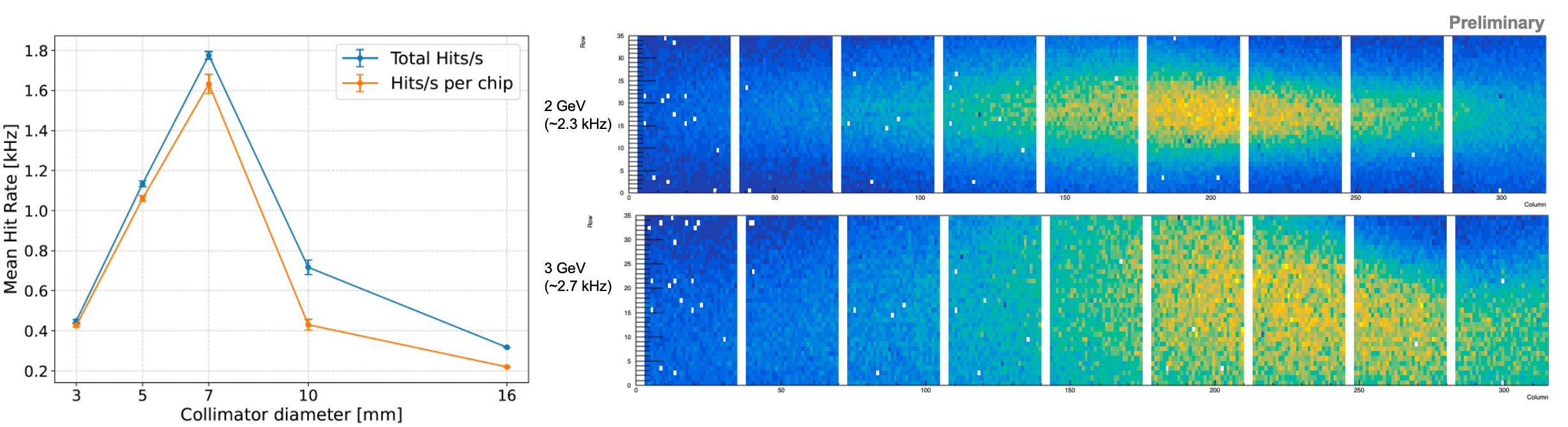}
    \caption{Mean hit rates per nine-chip module and per chip measured using a $^{90}$Sr source~\cite{kim2025astropix} (left), and beam profiles obtained with 2 GeV (top) and 3 GeV (bottom) electron beams (right).}
    \label{fig:v3nine}
\end{figure}

The AstroPix v3 nine-chip prototype was developed to validate the electrical design and readout architecture of the AstroPix module for the BIC, serving as a proof of concept for the module-level readout system. A data-rate study was performed to evaluate the readout capability of the nine-chip prototype~\cite{kim2025astropix}. As shown in the left plot of Figure~\ref{fig:v3nine}, the mean hit rate was measured as a function of collimator diameter, ranging from 3 to 16 mm, using a 10 $\mu$Ci $^{90}$Sr source. For reference, the measured per-chip hit rates were compared with the expected average and maximum hit rates in the BIC imaging layers of approximately 285 Hz and 1.18 kHz per chip, respectively, accounting for contributions from DIS events and electron- and proton-beam backgrounds. To verify operation under beam conditions, the nine-chip prototype was tested with electron beams at energies ranging from 1 to 5 GeV at KEK. Beam profiles obtained with 2 GeV (top) and 3 GeV (bottom) electron beams are shown in the right plots of Figure~\ref{fig:v3nine}. The first beam test demonstrated stable operation at beam rates of up to 2.7 kHz.

\section{Summary and Outlook}
The BIC combines mature Pb/SciFi sampling calorimeter technology with high-granularity AstroPix imaging layers to provide precise energy measurements, particle identification, and three-dimensional shower imaging for the ePIC detector. Prototype beam tests have demonstrated electron energy resolutions meeting the ePIC barrel EMCal system requirements, along with AstroPix MIP response, multi-chip operation, and multi-layer shower imaging capabilities. The BIC successfully completed its Preliminary Design Review (PDR) and Safety Review in FY25 and is progressing toward the Final Design Review (FDR).

In preparation for the FDR, BIC development follows a "learning by building" approach, progressing from small-scale prototypes to full-size detector components. Pb/SciFi development is advancing from 50 cm test articles through 3 m prototypes toward full-size sectors, while AstroPix imaging layer development progresses from nine-chip modules to half-stave and full-stave assemblies. Key prototypes, including Pb/SciFi calorimeter layers fabricated at Argonne, nine-chip AstroPix modules, and the first extruded-aluminum mechanical test article for the imaging staves, have been built and tested. 

In FY26, the first integration tests of the Pb/SciFi and AstroPix readout systems under beam conditions were conducted in Hall D at Jefferson Lab, and data analysis to evaluate event synchronization is currently ongoing.
Further system-level beam tests are planned for 2027 using a near-full-sector prototype comprising six Pb/SciFi layers and at least three multi-chip AstroPix imaging layers. These tests will focus on energy and timing resolutions, e/$\pi$ separation using AstroPix hit information, and validation of three-dimensional shower imaging performance. Pre-production preparations are also underway, including scintillating-fiber qualification for Pb/SciFi calorimeter production and quality control of AstroPix sensors.

\section*{Acknowledgment}
Development of AstroPix is funded through the following NASA programs: 18-APRA18-0084 and 20-RTF20-0003. This document was prepared by the members of the Barrel Imaging Calorimeter (BIC) Detector Subsystem Collaboration (DSC) of the ePIC Collaboration.
The material is based upon work partially supported by the U.S. Department of Energy, Office of Science, Office of Nuclear Physics and Laboratory Directed Research and Development (LDRD) funding from Argonne National Laboratory, provided by the Director, Office of Science, of the U.S. Department of Energy under Contract No.DE-AC02-06CH11357; the Electron-Ion Collider Project PED Funds.

\bibliographystyle{iopart-num}
\bibliography{bib.bib}

\end{document}